\documentclass[a4paper,fleqn,usenatbib]{mnras}

\usepackage[T1]{fontenc}
\usepackage{ae,aecompl}

\usepackage{graphicx}	
\usepackage{amsmath}	
\usepackage{amssymb}	

\title[RVM for magnetars]{Rotating vector model for magnetars}

\author[H. Tong et al.]{H. Tong$^{1}$\thanks{E-mail: htong\_2005@163.com},
P. F. Wang$^{2}$,
H. G. Wang$^{1}$,
Z. Yan$^{3}$
\\
$^{1}$School of Physics and Materials Science, Guangzhou University, Guangzhou 510006, China\\
$^{2}$National Astronomical Observatories, Chinese Academy of Sciences, A20 Datun Road, Chaoyang District, Beijing 100012, China\\
$^{3}$Shanghai Astronomical Observatory, Chinese Academy of Sciences, 80 Nandan Road, Shanghai 200030, China
}
\date{Accepted XXX. Received YYY; in original form ZZZ}

\pubyear{2015}

\begin{document}
\label{firstpage}
\pagerange{\pageref{firstpage}--\pageref{lastpage}}
\maketitle

\begin{abstract}
The modification of the rotating vector model in the case of magnetars are calculated. Magnetars may have twisted magnetic field compared with normal pulsars. The polarization position angle of magnetars will change in the case of a twisted magnetic field. For a twisted dipole field, we found that the position angle will change both vertically and horizontally. During the untwisting process of the magnetar magnetosphere, the modifications of the position angle will evolve with time monotonously. This may explain the evolution of the position angle in magnetar PSR J1622-4950 and XTE J1810-197. The relation between the emission point and the line of sight will also change. We suggest every magnetospheric models of magnetars also calculate the corresponding changes of position angle in their models. Order of magnitude estimation formula for doing this is given. This opens the possibility to extract the magnetic field geometry of magnetars from their radio polarization observations.
\end{abstract}



\section{Introduction}

Magnetars are thought to be young and strongly magnetized neutron stars (Kaspi \& Beloborodov 2017). There may be a continuous transition from normal pulsars to magnetars. Observational evidences include: (1) Five magnetars are observed to have pulsed radio emissions (Camilo et al. 2006, 2007a; Levin et al. 2010; Eatougth et al. 2013; Lower et al. 2020a). (2) Magnetar-like activities are observed in two high magnetic field pulsars and one is also radio-loud (Gavriil et al. 2008; Archibald et al. 2016). (3) Several low dipole magnetic field magnetars are also discovered (Rea et al. 2010; Zhou et al. 2014) etc. The burst and persistent X-ray emissions of magnetars are thought to be powered by the neuton star's strong magnetic field, i.e. the name ``magnetar'' (Duncan \& Thompson 1992). The reason why magnetar can have a huge magnetic energy release rate may be that their magnetic field is twisted compared with that of normal pulsars (Thompson et al. 2002; Beloborodov 2009; Paven et al. 2009; Tong 2019). The untwisting of the twisted magnetic field is responsible for the outburst activities seen in magnetars (Coti Zeltati et al. 2018).

Up to now, all the evidences for a twisted magnetic field in magnetars are indirect observations (Tiengo et al. 2013; Weng et al. 2015). Direct evidence of a twisted magnetic field may be revealed by future X-ray polarimetry observations. At the same time, for radio emitting magnetars, we already have polarization observations which may tell us the magnetic field geometry of magnetars (Camilo et al. 2007b; Kramer et al. 2007; Camilo et al. 2008; Levin et al. 2012; Eatought et al. 2013; Lower et al. 2020b). However, when analysing polarization position angle of the radio emitting magnetars, the rotating vector model for a dipole field is always employed (e.g., Levin et al. 2012; Lower et al. 2020b). The rotating vector model in the case of a twisted magnetic field is not available at present. This hinders us from extracting the magnetic field geometry information from magnetar radio observations.

We will build a rotating vector model in the case of twisted dipole field for magnetars. It is compared with the current magnetar radio observations. It is shown that the polarization of magnetar radio emission may be modeled to some degree by the twisted dipole field.

The structure of this paper is: the rotating vector model for a twisted dipole field is calculated using simple geometry in Section 2. A more detailed calculation is presented in Section 3. A twisted dipole field will also change the relation between the emission point and the line of sight. This is shown in Section 4. Comparison with the observations is presented in Section 5. Discussion and conclusion are given in Section 6. In normal pulsars, it is a common method to extract magnetic field geometry using the rotating vector model. The corresponding calculations in the dipole case are given in the Appendixes.

\section{Calculations using simple geometry}

\begin{figure}
\centering
\includegraphics[width=0.45\textwidth]{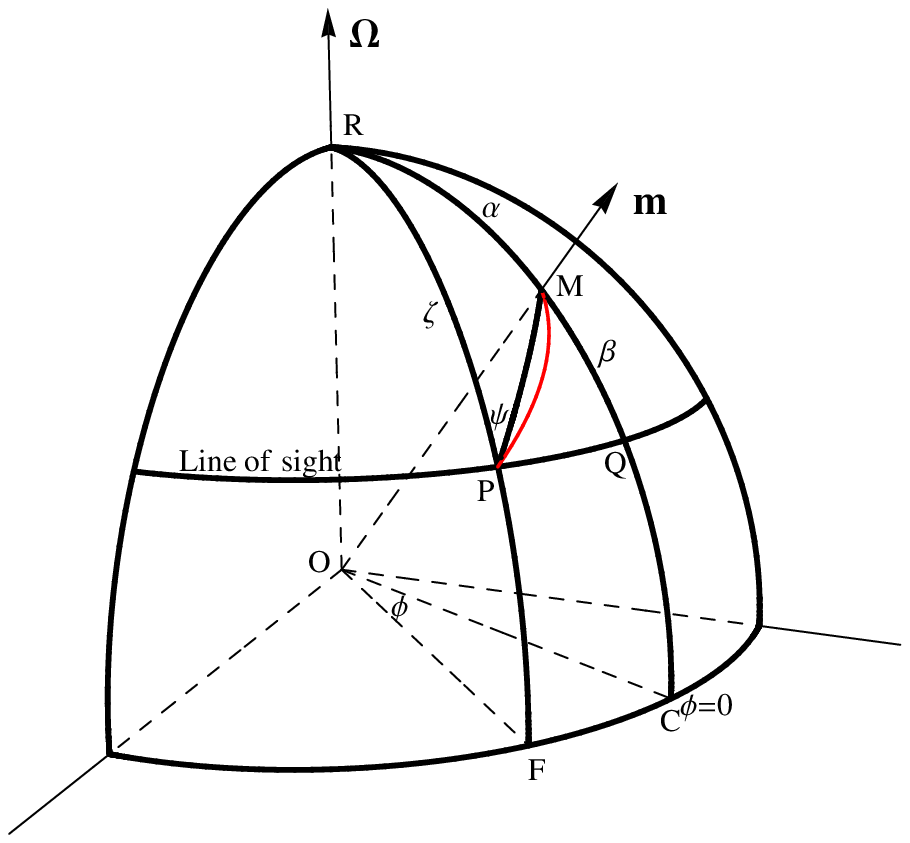}
\caption{Geometry of a rotating dipole. The rotational axis and the magnetic axis are shown, along with the line of sight.
At the observational point $P$, the magnetic field line makes a position angle $\psi$ with the rotational axis. The red solid line is for a twisted dipole field.}
\label{fig_gdipole}
\end{figure}

\begin{figure}
\centering
\includegraphics[width=0.45\textwidth]{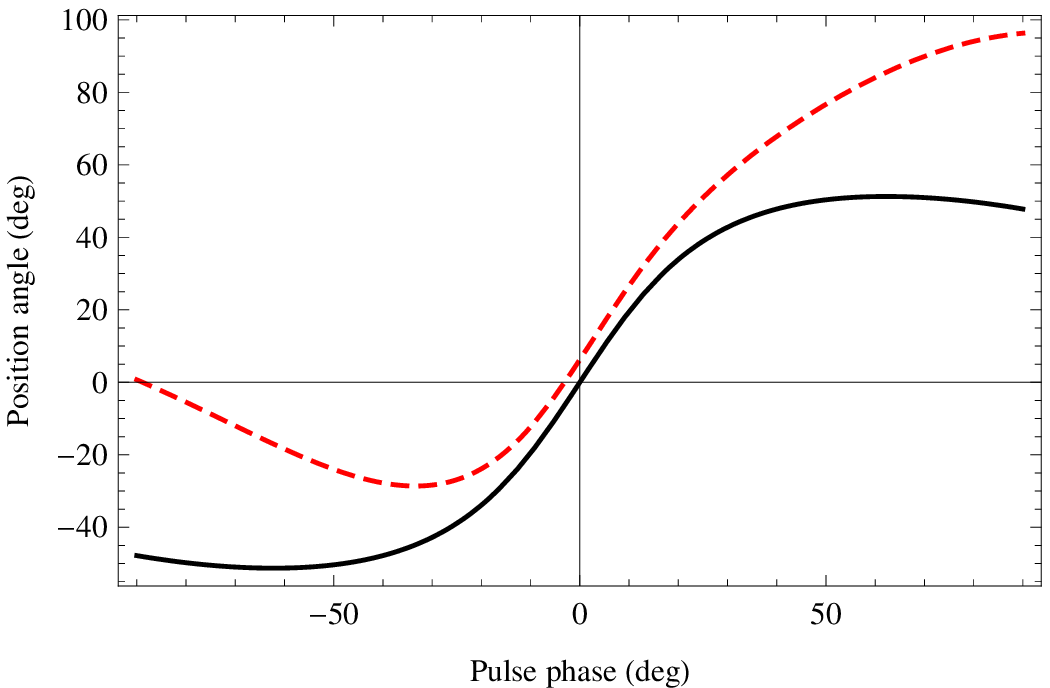}
\caption{Polarization position angle as a function of pulse phase. The black solid line is for a dipole magnetic field, with $\alpha =45^{\circ}$, $\zeta=65^{\circ}$ ($\beta = 20^{\circ}$). The red dashed line is for a twisted dipole field, with twist parameter $n=0.5$, the inclination angle and viewing angle is the same as the dipole case. }
\label{fig_gPA}
\end{figure}

The geometry of a rotating dipole is shown in figure \ref{fig_gdipole}. The angle between the magnetic axis and the rotational axis is denoted as $\alpha$ (inclination angle), the angle between the line of sight and the rotational axis is denoted as $\zeta$ (viewing angle). The plane where the rotational axis and the magnetic axis lies is denoted as $\phi=0$ (the meridian plane). The closest approach between the line of sight and magnetic axis is denoted as $\beta$ (impact angle). From figure \ref{fig_gdipole}, it can seen that $\beta = \zeta -\alpha$.  The position angle $\angle MPR= \psi$ can be found as a function of the rotational phase $\angle MRP = \phi$
\begin{equation}\label{eqn_psidipole}
  \tan\psi = \frac{\sin\alpha \sin\phi}{\cos\alpha \sin\zeta -\sin\alpha \cos\zeta \cos\phi}.
\end{equation}
The deduction process is given in Appendix \ref{appendixA}. Polarization position angle curve for typical parameters is shown in figure \ref{fig_gPA}.
A typical inclination angle ($\alpha=45^{\circ}$) and a large impact angle ($\beta=20^{\circ}$) is chosen in figure \ref{fig_gPA}. Please note that the position angle defined in equation (\ref{eqn_psidipole}) and the observationally reported position angle is different by a minus sign: $\psi_{\rm obs} = -\psi$ (Everett \& Weisberg 2001). This will account for the position angle of PSR J1622-4950: it has a positive slope in figure 4 in Levin et al. (2012), while the reported impact angle $\beta$ is negative in Levin et al. (2012).

\begin{figure}
\centering
\includegraphics[width=0.45\textwidth]{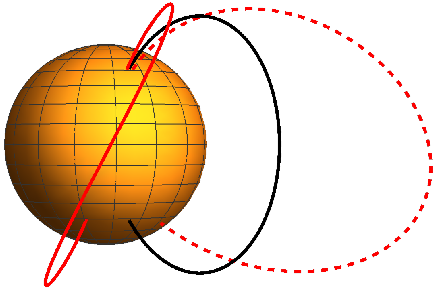}
\caption{Twisted magnetic field lines. The black solid line is for a dipole field. The red solid line is for a twisted dipole field. It twists westward in the southern hemisphere. The red dashed line twists eastward in the southern hemisphere.}
\label{fig_twist}
\end{figure}

For a twisted dipole field, the magnetic field will have a toroidal component compared with the dipole case (Wolfson 1995; Thompson et al. 2002; Pavan et al. 2009; Tong 2019). The toroidal component will result in modifications to the position angle. For small twist, the magnetic field can be considered as a dipole field (which is purely poloidal) plus the toroidal component (Beloborodov 2009; Tong 2019). Therefore, the corresponding position angle may also be considered as the dipole case plus some modification
\begin{equation}\label{eqn_psi_total}
  \psi = \psi (\rm dipole) + \Delta \psi_{\rm twist},
\end{equation}
where $\psi (\rm dipole)$ is the corresponding value given in equation (\ref{eqn_psidipole}), and $\Delta \psi_{\rm twist}$ is due to the twist of magnetic field lines. For a self-similar twisted dipole field with radial dependence $B(r) \propto r^{-(2+n)}$, the twist angle will be the same for the same initial and final magnetic colatitude. Here the parameter $n$ characterize the twist of the magnetic field: $n=1$ corresponds to the dipole case, $n=0$ the split monopole case, and $n\in(0,1)$ is the general twisted dipole case.  When referring to the magnetic field, we will mainly work in the magnetic frame, where the magnetic axis is the spherical polar axis. For an initial magnetic colatitude $x_1 = \cos\theta_1$, final magnetic colatitude $x_2 =\cos\theta_2$, the twist of magnetic field is (Wolfson 1995)
\begin{equation}
  \Delta \phi =\frac{\lambda}{n} \int_{x_2}^{x_1} \frac{f^{1/n}(x)}{1-x^2} dx,
\end{equation}
where
\begin{equation}\label{eqn_lambda}
\lambda=\sqrt{(35/16)(1-n)},
\end{equation}
and $f(x)$ is the dimensionless flux function. About the twist direction of the magnetic field, a positive $\lambda$ corresponds to eastward motion of magnetic field lines in the southern hemisphere (Tong 2019). See figure \ref{fig_twist} for an example. The twist field lines are also shown in figure 2 in Wolfson (1995) and figure 3 in Thompson et al. (2002). The modification to the position angle is positive when the field line twist westward (as shown in figure \ref{fig_gdipole}, negative $\lambda$).  Otherwise, it will be negative (positive $\lambda$).

The maximum twist is defined as the twist along the magnetic field line from the north pole to the sourth pole
\begin{eqnarray}
  \Delta \phi_{\max} &=& \frac{\lambda}{n} \int_{-1}^{1} \frac{f^{1/n}(x)}{1-x^2} dx\\
  &=& \frac{2\lambda}{n} \int_{0}^{1} \frac{f^{1/n}(x)}{1-x^2} dx\\
  \label{eqn_delta_phi_max}
  &\approx & 2\lambda,
\end{eqnarray}
where during the calculations, the small twist approximation is used $n\approx 1$, and $f(x) \approx 1-x^2$.
This approximation is accurate for a wide parameter space (Tong 2019). Therefore, there is a one to one corresponds between the three paprameters: $n$, $\lambda$, and $\Delta \phi_{\rm max}$. For the case shown in figure \ref{fig_gdipole},
the modification to the position angle is
\begin{eqnarray}
  \Delta \psi_{\rm twist} &=& \frac{\lambda}{n} \int_{\theta_{\rm obs}}^1 \frac{f^{1/n}(x)}{1-x^2} dx\\
  & \approx & \frac{\lambda}{n} (1-\cos\theta_{\rm obs})\\
  & \approx & \frac{\lambda}{2n} \sin^2\theta_{\rm obs},
\end{eqnarray}
where during the last step, the approximation of small magnetic colatitude is used $1-\cos\theta_{\rm obs} \approx 1/2 \sin^2\theta_{\rm obs}$. Adding the correct sign of the twist angle, the final result is
\begin{eqnarray}
  \Delta \psi_{\rm twist} &=& -\frac{\lambda}{n} (1-\cos\theta_{\rm obs}) \\
  \label{eqn_RVM_twist}
  &\approx& -\frac{\lambda}{2n} \sin^2\theta_{\rm obs}.
\end{eqnarray}
As discussed above, a positive $\lambda$ corresponds to a magnetic field twist eastward. The position angle will be smaller.
This corresponds to the negative sign in the equation (\ref{eqn_RVM_twist}).
In below, equation (\ref{eqn_RVM_twist}) is confirmed using more detailed calculations.

From equation (\ref{eqn_RVM_twist}), it can be seen that the modification to the position angle
\begin{equation}\label{eqn_psi_decay}
\Delta \psi_{\rm twist} \propto \lambda \propto \sqrt{1-n} \propto e^{-t/2\tau},
\end{equation}
where during the last step the evolution of the parameter $n$ is adopted from the toy model for magnetar outburst (Tong \& Huang 2020). The parameter $\tau$ is the exponential decay timescale of the X-ray flux or hot spot. Therefore, the modification to the position angle, especially the constants of the position angle $\psi_0$, will evolve with time monotonously. This is consistent with the polarization observation of the radio emitting magnetar PSR J1622$-$4950 (Levin et al. 2012). Quantitative comparison will be given in below.

\section{More detailed calculations}

The above calculations may seem too simple minded. A more detailed calculation can be done following the procedure of Hibschman \& Arons (2001). The details and corresponding calculations for the dipole case are given in Appendix \ref{appendixB}. Considering the rotation of the neutron star (Blaskiewicz et al. 1991), or the magnetospheric current (Hibschman \& Arons 2001), the change of position angle may be obtained using perturbations. These two effects are both proportional to $r/R_{\rm L}$, where $r$ is the radio emission height, and $R_{\rm L}$ is the light cylinder radius. For a twisted magnetic field, the result will be qualitatively different. In the case of a self-similar twisted dipole field, the modification to the position angle is independent of the emission height. This is because all the field component (poloidal and toroidal) have the same radial dependence $\propto r^{-(2+n)}$ (Wolfson 1995).

For an axisymmetric force-free magnetic field, the magnetic field obeys the Grad-Shafranov equation. In the magnetic frame, using spherical coordinate, the magnetic field is (Wolfson 1995)\footnote{A normalization factor for the magnetic field omitted. It can be normalized to the polar or equatorial surface magnetic field. Only the geometry of the magnetic field is included.}
\begin{equation}
\label{eqn_twisted_field}
  {\bf B} = \frac{1}{r\sin\theta} [ \frac{1}{r} \frac{\partial A}{\partial \theta} \hat{r} - \frac{\partial A}{\partial r} \hat{\theta}
  + F(A) \hat{\phi} ].
\end{equation}
For the self-similar case, the two functions are: $F(A) = \lambda A^{1+1/n}$, $A(x) = r^{-n} f(x)$ (where $x=\cos\theta$).
The toroidal magnetic field component may be considered as perturbation to the dipole magnetic field (Beloborodov 2009; Tong 2019).
Therefore, the perturbation to the tangent vector is
\begin{eqnarray}
  {\bf t}_1 &=& \frac{B_{\phi}}{|\bf B|} \hat{\phi}
  \approx \frac{B_{\phi}}{|{\bf B}_{\rm p}|} \hat{\phi}
  = \frac{B_{\phi}}{(B_r^2 + B_{\theta}^2)^{1/2}} \hat{\phi}\\
  &=& \frac{F(A)}{[(1/r \, \partial A/\partial\theta)^2 + (\partial A/\partial r)^2]^{1/2}} \hat{\phi}\\
  \label{eqn_t1_complex}
  &=& \lambda \frac{f^{1+1/n}(x)}{[(\partial f/\partial \theta)^2 + n^2 f^2(x)]^{1/2}} \hat{\phi}.
\end{eqnarray}
In order to an meaningful analytical result, further simplifications are needed. At the same time, it is required that when the parameter $n$ is back to one, the result should be back to the dipole case. Taken the flux function as the dipole case $f(x) \approx 1-x^2 = \sin^2\theta$, the parameter $n$ in the power law index of $f(x)$ set to be $n=1$, while keeping the expression of $\lambda$ unchanged, equation (\ref{eqn_t1_complex}) can be further simplified
\begin{eqnarray}
  {\bf t}_1 &=& \lambda \frac{\sin^3\theta}{[4\cos^2\theta + \sin^2 \theta]^{1/2}} \hat{\phi} \\
  &\approx& \frac{\lambda}{2} \sin^3 \theta \hat{\phi},
\end{eqnarray}
where during the last step the approximation of small $\theta$ is used.

Following the procedure outline in Appendix \ref{appendixB},
it is straight forward (though tedius) to calculate the changes in the position angle due to a twisted dipole field
\begin{eqnarray}
  \Delta \psi_{\rm twist} &=& -2 \lambda \sin^2\theta \\
  \label{eqn_RVM_twist_detailed_calculation}
  &\approx & -\frac{8}{9} \lambda \sin^2\theta_{\rm obs},
\end{eqnarray}
where during the last step the relation between the line of sight and the emission point are used: $\theta \approx 2/3 \, \theta_{\rm obs}$, $\theta$ is the colatitude of the emission point, $\theta_{\rm obs}$ is the colatitude of the line of sight, both in the magnetic frame. $\theta_{\rm obs}$ can be found as a function of the pulsar rotational phase, see equation (\ref{eqn_thetaobs}).  For positive $\lambda$, the magnetic field will twist eastward. This will result in a smaller position angle. This corresponds to the minus sign in equation (\ref{eqn_RVM_twist_detailed_calculation}). The result in equation (\ref{eqn_RVM_twist_detailed_calculation}) is consistent with that in equation (\ref{eqn_RVM_twist}). The observational meaning of equation (\ref{eqn_RVM_twist_detailed_calculation}) is that a twisted dipole field will change the polarization position angle in the vertical direction. Figure \ref{fig_gPA} shows the modified position angle for a twisted dipole field, for twist parameter $n=0.5$. A relatively large impact parameter is chosen. A large impact angle will result in larger deviation from the dipole case. For a self-similar twisted dipole field, there is one additional parameter $n$ (or $\lambda$, or $\Delta \phi_{\rm max}$, there is a one to one correspondence between these three parameters) compared with the dipole case. This is the merit of assuming a self-similar twisted dipole field.

The relation between the emission point and the line of sight ($\theta \approx 2/3 \, \theta_{\rm obs}$) is for the dipole case (Hibschman \& Arons 2001). For a twisted dipole field, the relation between the emission point and the line of sight is only slightly different, as detailed in below.

\section{Relation between emission point and line of sight}

The relation between the emission point and the line of sight for a dipole field is shown in Appendix \ref{appendixC}.
For a twisted dipole field, the tangent vector can be get from equation (\ref{eqn_twisted_field})
\begin{equation}
  {\bf \hat{t}} = \left( \frac{1}{\sin\theta} \frac{\partial f}{\partial \theta} \hat{r} + n \frac{1}{\sin\theta} f(x) \hat{\theta} +
  \lambda \frac{1}{\sin\theta} f^{1+1/n}(x) \hat{\phi} \right) \frac{1}{N},
\end{equation}
where $N$ is the normalization factor. For the small twist case, the expression for the tangent vector can be further simplified
\begin{equation}
  {\bf \hat{t}} = \left( 2\cos\theta \hat{r} + \sin\theta \hat{\theta} + \lambda \sin^3\theta \hat{\phi} \right) \frac{1}{N}.
\end{equation}
In the small twist case, the normalization factor is $N = (4\cos^2\theta + \sin^2\theta + \lambda^2 \sin^6\theta)^{1/2} \approx (4-3\sin^2\theta)^{1/2}$. In comparison with the normalization factor defined in Appendix \ref{appendixC}, here the normalization factor is $N \approx 2 N_1$. The main deviation from the dipole case is a toroidal magnetic field component. The tangent vector can be expressed in Cartesian coordinate. By equaling the tangent vector and the unit vector along the line of sight, it is found that
\begin{eqnarray}
\label{eqn_direction_1}
  \frac{1}{N} (3\sin\theta \cos\theta \cos\phi - \lambda \sin^3\theta \sin\phi) = \sin\theta_{\rm obs} \cos\phi_{\rm obs} &&\\
\label{eqn_direction_2}
  \frac{1}{N} (3\sin\theta \cos\theta \sin\phi + \lambda \sin^3\theta \cos\phi) = \sin\theta_{\rm obs} \sin\phi_{\rm obs} &&\\
\label{eqn_direction_3}
  \frac{1}{N} (2\cos^2\theta -\sin^2\theta) = \cos\theta_{\rm obs}.&&
\end{eqnarray}
From equation (\ref{eqn_direction_3}), it can be inferred that the changes in $\theta$ is only quantitative. The relation between $\theta$ and $\theta_{\rm obs}$ may be taken as the same as the dipole case $\theta = (2/3) \theta_{\rm obs}$. The major changes is for the azimuthal angle: $\phi = \phi_{\rm obs} + \delta \phi$. By substituting into equation (\ref{eqn_direction_2}), keeping only the first order term, it is found that
\begin{equation}\label{eqn_delta_phi}
  \delta \phi = -\frac{4}{27} \lambda \sin^2\theta_{\rm obs}.
\end{equation}
For a positive $\lambda$, the field will twist eastward. The azimuthal angle $\phi$ will be smaller than $\phi_{\rm obs}$. Therefore, the minus sign in the above equation is correct. Equation (\ref{eqn_direction_1}) will give the same result. The observational meaning of equation (\ref{eqn_delta_phi}) is that a twisted dipole field will change the polarization position angle in the horizontal direction.

By comparing equation (\ref{eqn_RVM_twist_detailed_calculation}) and equation (\ref{eqn_delta_phi}) it can be seen that a twisted dipole field will change the polarization position angle in both the vertical and the horizontal direction. Furthermore, the changes in the vertical and the horizontal direction have the same dependence on the twist parameter $\lambda$ and the colatitude of the line of sight in the magnetic frame $\theta_{\rm obs}$. Quantitatively, the change in the horizontal direction is only $1/6$ of that in the vertical direction $\delta \phi =1/6 \ \Delta \psi_{\rm twist}$. Therefore, the change of position angle in the vertical direction may be easier to observe.

\section{Comparison with the observations}

The rotating vector model is commonly used in the case of normal pulsars to obtain the information of the magnetic field geometry (Lyne \& Manchester 1988; Rankin 1993; Manchester et al. 1998; Johnston \& Weisberg 2006; Johnston \& Kramer 2019). The position angle of some pulsars may be too complicated to be modeled by the rotating vector model.

\begin{figure}
\centering
\includegraphics[width=0.45\textwidth]{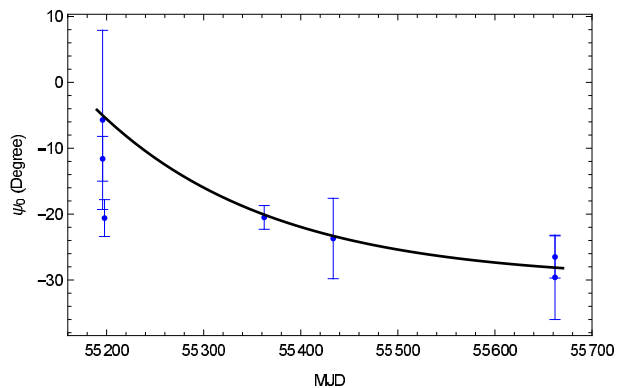}
\caption{The evolution of $\psi_0$ of the radio emitting magnetar PSR J1622-4950. The data points are from table 1 of Levin et al. (2012) (one data point with position angle jump is not considered). The solid line is a typical curve of exponential decay form: $\psi_0(t) = -30 + 25\exp(-(t-t_0)/2\tau)$ (in unit of degrees), where $t_0$ is the date of the first data points, and the exponential decay timescale is $\tau= 90$ days. }
\label{fig_g1622}
\end{figure}

In the case of magnetars, previous polarization observations of five radio emitting sources may already tell us some aspect of the magnetic field geometry of magnetars. In our opinion, current observations already give us some hints that the magnetic field of magnetars may be twisted. And the rotating vector model for a twisted dipole field may explain the observations to some degree.
\begin{enumerate}
  \item Changing $\psi_0$. For the third radio emitting magnetar PSR J1622-4950, its position angle changes monotonously with time (figure 4 and table 1 in Levin et al. 2012). For a twisted dipole field, the position angle will change in the vertical direction (see equation (\ref{eqn_RVM_twist_detailed_calculation})). This is also true for the position angle in the meridian plane $\psi_0$. During the untwisting process, the magnetic field will relax to that of the dipole case (Thompson et al. 2002; Tong 2019). The parameter $n$ will approach 1. The parameter $\lambda$ will decrease monotonously with time (equation (\ref{eqn_lambda})). According to the toy model for magnetar outburst (Tong \& Huang 2020), the parameter $\lambda$ will decrease with time exponentially. The corresponding position angle, also for $\psi_0$, will decrease with time exponentially (equation(\ref{eqn_psi_decay})). This is consistent with the observations of PSR J1622-4950 (Levin et al. 2012). Figure \ref{fig_g1622} shows quantitatively that the $\psi_0$ of PSR J1622-4950 may decrease in an exponential form\footnote{Mathematically, the observational data may also be fitted using other forms, e.g. a linear decay form etc. According to our physical modeling, we prefer an exponential decay form.}.
      A decreasing $\psi_0$ is also seen in the fifth radio emitting magnetar Swift J1818.0-1607 (Lower et al. 2020b, table 3 there).
      Again, this indicates an untwisting magnetic field in the putative magnetar.

      From the typical exponential decay form in figure \ref{fig_g1622}, the decay time scale is about $90$ days. This is smaller than the X-ray flux decay time scale (360 days, Anderson et al. 2012). But they can be considered as the same order of magnitude. The typical decay amplitude is about $25^{\circ}$. According to equation (\ref{eqn_RVM_twist_detailed_calculation}), the changes of position angle at the meridian plane is about: $|\Delta \psi_0|= (8/9) \lambda \sin^2\beta$, where $\beta$ is the impact angle. For typical twist parameter $n=0.5$, the required impact angle for PSR J1622-4950 in the twisted dipole scenario is about: $|\beta| = 40^{\circ}$. Observationally, the range of impact angle for PSR J1622-4950 is: $\beta \in [-25^{\circ}, -8^{\circ}]$ (Levin et al. 2012). The theoretical value of the impact angle $\beta$ is slightly larger than the observational values. But again, they can be considered as roughly consistent. In the presence of multipole field, the modification to the twist angle may be larger.

  \item Changing slope. For the radio emitting magnetar PSR J1622-4950, the slope of the position angle is also changing with time (Section 5.3 in Levin et al. 2012). A changing slope of the position angle is also seen in the first radio emitting magnetar XTE J1810-197 (Section 3.4 in Kramer et al. 2007). During the untwisting process of the magnetar's magnetic field (Thompson et al. 2002; Tong 2019), the parameter $n$, $\lambda$ etc are all evolve with time monotonously. Then the change in position angle due to the magnetic field twist will also evolve with time. This may explain the changing slope of position angle in XTE J1810-197 and PSR J1622-4950.

  \item Changing $\phi_0$. During the revival of radio emission in PSR J1622-4950 (section 4.2 in Camilo et al. 2018), the deflection point of the position angle is changed horizontally. The overall shape of the position angle changes from convex to concave. After the magnetar-like activities in the high magnetic field pulsar PSR J1119-6127, the slope of the position angle even changes sign (Dai et al. 2018). In our opinion, the twist of the magnetic field can change the position angle horizontally (equation (\ref{eqn_delta_phi})). This may explain the change in deflection point in PSR J1622-4950. The problem is that the changes in the horizontal direction for a twisted dipole field is relatively small (see discussions following equation (\ref{eqn_delta_phi})). If the field geometry is more complicated than the twisted dipole case, large changes in the horizontal direction may be possible.

      Furthermore, if there are large changes in the position angle horizontally, the overall shape of the position angle may change from convex to concave as can be seen in figure \ref{fig_gPA}. This may correspond to the case of PSR J1622-4950 (Camilo et al. 2018).
      If the horizontal change of position angle is large enough, the slope of the position angle can even change sign. This may explain the observations of PSR J1119-6127 (Dai et al. 2018). For large changes in the horizontal direction, a more complicated field geometry than the twisted dipole field is required.

  \item Wide pulse profile. When the rotating vector model is proposed (Radhakrishnan \& Cooke 1969), it provides an explanation for both the pulse profile and position angle. The five radio emitting magnetars all have a period about several seconds. However, their radio pulse profile are very wide, for a duty cycle about $50\%$. This is much wider than that of normal pulsars (Lyne \& Manchester 1988; Tan et al. 2018). One reason may be that the radio emission of magnetars originates from a much higher radius in the magnetosphere. Considering that the magnetars' magnetic field is twisted, it is more probable that the polar cap will be larger for a twisted dipole field (Tong 2019).  During untwisting process, the polar cap will be smaller. This may explain the shrinking hot spot of magnetar X-ray observations (Tong 2019). This may also explain the wide pulse profile of magnetar radio emissions. One prediction of the twisted dipole field explanation is that: during the untwisting process, the pulse profile will become narrower. This is indeed observed during the diminishing process PSR J1622-4950 (Scholz et al. 2017).
\end{enumerate}

In summary, the ordered position angle of magnetar radio emission tells us that the large scale magnetic field in the radio emitting region is ordered, e.g., dipole or twisted dipole. The time varying position angle (change $\psi_0$, slope, $\phi_0$) indicates time evolution of the magnetosphere. A twisted dipole field and its untwisting process may explain the polarization of magnetar radio emissions. A wide pulse profile and its evolution are also consistent with a twisted dipole field. More complex field geometry is needed in order to explain the changes in $\phi_0$ quantitatively.

\section{Discussion and conclusion}

In order to have high quality polarization measurement, a high flux of the source is preferred. This means that the polarization measurement of magnetar is more easily obtained at the beginning of the radio observations (Eatough et al. 2013). This may also explain why only XTE J1810-197 and PSR J1622-4950 have detailed polarization information (Camilo et al. 2007; Kramer et al. 2007; Dai et al. 2019; Levin et al. 2012; Camilo et al. 2018). The other three radio emitting magnetars also have polarization observations, but the details are limited compared with XTE J1810-197 and PSR J1622-4950. Therefore, more observations are required in order to infer the magnetic field geometry from radio polarization data.

Considering that there is a continuous transition from normal pulsars to magnetars, twist of magnetic field may also happen in normal pulsars. This possibility is strengthened by the two high magnetic field pulsars with magnetar activities (Gavriil et al. 2008; Archibald et al. 2016). Weak and strong twist may account for different behaviors of high magnetic field pulsars and magnetars (Tong 2019). The twist in the case of normal pulsars may be very weak. And there exist several depolarization effects (Levin et al. 2012). Therefore, the effect of a twisted magnetic field in normal pulsar may be very difficult to verify. However, with the FAST or SKA telescope, we may find hints of a twisted magnetic field in some normal pulsars.

In principle, we should fit the observed position angle using the exact formula (equation (\ref{eqn_psi_total})): the pure dipole case given by equation (\ref{eqn_psidipole}), the modification in the twisted dipole case given by equation (\ref{eqn_RVM_twist_detailed_calculation}), the twist parameter $\lambda$ can be either positive or negative, and $\theta_{\rm obs}$ is also determined by the geometry parameters (equation (\ref{eqn_thetaobs})). The twisted dipole scenario only introduce one additional free parameter $\lambda$ (or $n$, or $\Delta \phi_{\rm max}$ which are equivallent). From previous examples of PSR J1622-4950, if a changing $\psi_0$ or $\phi_0$ or changing slope etc are seen in one radio emitting magnetar, it may due to an untwisting magnetosphere. Irrespective of this point, the observed position angle can also be fitted by the dipole formula. Then the time evolving $\psi_0$ can be modeled by the twisted dipole field etc. This is the second way of fitting the observations. The merit of the second method is that the changing $\psi_0$ can be modeled by the twisted dipole field or other magnetospheric models of magnetars.

When modeling the X-ray outburst of magnetars, the modeling is often based on a twisted magnetic field (Thompson et al. 2002; Beloborodov 2009; Pavan et al. 2009; Fujisawa \& Kisaka 2014; Akgun et al. 2016; Kojima 2017; Tong 2019). We suggest that the researchers also calculate the corresponding polarization position angle in their magnetic field geometry. The procedure can be similar to that in Section 2 (simple geometry calculation) or Section 3 (more detailed calculations based on a perturbation analysis). For any magnetic field geometry, an order of magnitude estimation can be done. According to the definition of the magnetic field lines (Wolfson 1995)
\begin{equation}
  \frac{r\sin\theta d\phi}{r d\theta} = \frac{B_{\phi}}{B_{\theta}}.
\end{equation}
Therefore, in order of magnitude, the changes of the field line in the $\phi$-direction is
\begin{equation}
  \Delta \phi_{\rm twist} \sim \frac{B_{\phi}}{\sin\theta B_{\theta}} \sim \frac{B_{\phi}}{\sin\theta B_{\rm p}},
\end{equation}
where $B_{\phi}$ is the toroidal magnetic field, and $B_{\rm p}$ is the poloidal magnetic field. Adding the correct sign, the change in the position angle is
\begin{equation}\label{eqn_RVM_estimation}
  \Delta \psi_{\rm twist} \sim - 4 \frac{B_{\phi}}{\sin\theta B_{\rm p}}.
\end{equation}
The numerical factor $4$ is obtained by comparison with the detailed calculations in the twisted dipole case (equation (\ref{eqn_RVM_twist_detailed_calculation})). Therefore, for any magnetospheric modeling of magnetars, the corresponding modifications to the position angle can be estimated using equation (\ref{eqn_RVM_estimation}). Only the relative strength between the toroidal field and the poloidal field is required.

If the toroidal field is due to the polar cap current (Hibschman \& Arons 2001), according to equation (\ref{eqn_RVM_estimation}), an order of magnitude estimation is:  $\Delta \psi_{\rm twist} \sim 4 (r/R_{\rm L}) (J/J_{\rm GJ}) \cos\alpha$. Comparing the estimation here and equation (D13) in Hibschman \& Arons (2001), it can be seen that equation (\ref{eqn_RVM_estimation}) can give the correct sign and correct order of magnitude estimation for small $\theta$.

In conclusion, the twisted magnetic field of magnetars will bring changes to their polarization position angle. The relation between the emission point and the line of sight is slightly modified. Current radio observations of magnetars may indicate that their magnetic field is indeed twisted. At present, only the twisted dipole case is calculated. Calculations for more complicated field geometry (e.g., multipole field) are needed in the future.

\section*{Acknowledgements}

H.Tong is supported the Ministry of Science and Technology of the People's Republic of China (2020SKA0120300) and NSFC (11773008).
The authors would like to thank R.X.Xu for helpful comments.

\section*{Data availability}

This is a theoretical paper, mainly analytical. All the formulae are available in the article.




\begin{thebibliography}{99}

\bibitem{Akgun2016}
Akg\"{u}n T., Miralles J. A., Pons J. A., et al., 2016, MNRAS, 462, 1894

\bibitem{Anderson2012}
Anderson G. E., Gaensler B. M., Slane P. O., et al., 2012, ApJ, 751, 53

\bibitem{Archibald2016}
Archibald R. F., Kaspi V. M., Tendulkar S. P., et al., 2016, ApJ, 829, L21

\bibitem{Beloborodov2009}
Beloborodov A. M., 2009, ApJ, 703, 1044

\bibitem{Blaskiweicz1991}
Blaskiewicz M., Cordes J. M., Wasserman I., 1991, ApJ, 370, 643

\bibitem{Camilo2006}
Camilo F., Ransom S. M., Halpern J. P., et al., 2006, Nature, 442, 892

\bibitem{Camilo2007a}
Camilo F., Ransom S. M., Halpern J. P., et al., 2007a, ApJ, 666, L93

\bibitem{Camilo2007b}
Camilo F., Reynolds J., Johnston S., et al., 2007b, ApJ, 659, L37

\bibitem{Camilo2008}
Camilo F., Reynolds J., Johnson S., et al., 2008, ApJ, 679, 681

\bibitem{Camilo2016}
Camilo F., Ransom S. M., Halpern J. P., et al., 2016, ApJ, 820, 110

\bibitem{Camilo2018}
Camilo F., Scholz P., Serylak M., et al., 2018, ApJ, 856, 180

\bibitem{Carroll2018}
Carroll B. W., Ostlie D. A., 2018, An introduction to modern astrophysics, Cambridge, Cambridge

\bibitem{CotiZelati2018}
Coti Zelati F., Rea N., Pons J. A., et al., 2018, MNRAS, 474, 961

\bibitem{Dai2018}
Dai S., Johnson S., Weltevrede P., et al., 2018, MNRAS, 480, 3584

\bibitem{Dai2019}
Dai S., Lower M. E., Bailes M., et al., 2019, ApJL, 874, L14

\bibitem{DT1992}
Duncan R. C., Thompson C., 1992, ApJL, 392, L9

\bibitem{Eatough2013}
Eatough R. P., Falcke H., Karuppusamy R., et al., 2013, Nature, 501, 391

\bibitem{Everett2001}
Everett J. E., Weisberg J. M., 2001, ApJ, 553, 341

\bibitem{Fujisawa2014}
Fujisawa K., Kisaka S., 2014, MNRAS, 445, 2777

\bibitem{Gavriil2008}
Gavriil F. P., Gonzalez M. E., Gotthelf E. V., et al., 2008, Science, 319, 1802

\bibitem{Griffiths2017}
Griffiths D. J., 2017, Introduction to electrodynamics, Cambridge, Cambridge

\bibitem{Hibschman2001}
Hibschman J. A., Arons J., 2001, ApJ, 546, 382

\bibitem{Johnston2006}
Johnston S., Weisberg J. M., 2006, MNRAS, 368, 1856

\bibitem{Johnston2019}
Johnston S., Kramer M., 2019, MNRAS, 490, 4565

\bibitem{Kaspi2017}
Kaspi V. M., Beloborodov A. M., 2017, ARAA, 55, 261	

\bibitem{Kojima2017}
Kojima Y., 2017, MNRAS, 468, 2011

\bibitem{Kramer2007}
Kramer M., Stappers B. W., Jessner A., et al., 2007, MNRAS, 377, 107

\bibitem{Levin2010}
Levin L., Bailes M., Bates S., et al., 2010, ApJ, 721, L33

\bibitem{Levin2012}
Levin L., Bailes M., Bates S. D., et al., 2012, MNRAS, 422, 2489

\bibitem{Lower2020a}
Lower M. E., Shannon R. M., Johnson S., Bailes M., 2020a, ApJL, 896, L37

\bibitem{Lower2020b}
Lower M. E., Johnston S., Shannon R. M., Bailes M., Camilo F., 2020b, arXiv:2011.12463

\bibitem{Lyne1988}
Lyne A. G., Manchester R. N., 1988, MNRAS, 234, 477

\bibitem{Manchester1998}
Manchester R. N., Han J. L., Qiao G. J., 1998, MNRAS, 295, 280

\bibitem{Pavan2009}
Pavan L., Turolla R., Zane S., et al., 2009, MNRAS, 395, 753

\bibitem{RC1969}
Radhakrishnan V., Cooke D. J., 1969, ApL, 3, 225	

\bibitem{Rankin1993}
Rankin J. M., 1993, ApJ, 405, 285

\bibitem{Rea2010}
Rea N., Esposito P., Turolla R., et al., 2010, Science, 330, 944

\bibitem{Scholz2017}
Scholz P., Camilo F., Sarkissian J., et al., 2017, ApJ, 841, 126

\bibitem{Tan2018}
Tan C. M., Bassa C. G., Cooper S., et al., 2018, ApJ, 866, 54

\bibitem{Thompson2002}
Thompson C., Lyutikov M., Kulkarni S. R., 2002, ApJ, 574, 332

\bibitem{Tiengo2013}
Tiengo A., Esposito P., Mereghetti S., et al., 2013, Nature, 500, 312

\bibitem{Tong2019}
Tong H., 2019, MNRAS, 489, 3769

\bibitem{Tong2020}
Tong H., Huang L., 2020, MNRAS, 497, 2680

\bibitem{Wolfson1995}
Wolfson R., 1995, ApJ, 443, 810

\bibitem{Weng2015}
Weng S. S., Gogus E., Guver T., Lin L., 2015, ApJ, 805, 81

\bibitem{Zhou2014}
Zhou P., Chen Y., Li X. D., et al., 2014, ApJ, 781, L16


\end{thebibliography}



\appendix

\section{Rotating vector model for a dipole field}
\label{appendixA}

The expression of the  position angle for a rotating dipole can be found using spherical trigonometry.
From figure \ref{fig_gdipole}, the spherical triangle $\triangle RMP$ has two sides and one angle known: $\overline{RM}=\alpha$, $\overline{RP}=\zeta$, and $\angle MRP = \phi$.
The third side $\overline{PM}=\theta_{\rm obs}$ is the colatitude of the line of sight in the magnetic frame. It can be found using spherical trigonometry (Carroll \& Ostlie 2018). Using the law of cosines for sides
\begin{equation}\label{eqn_thetaobs}
  \cos\theta_{\rm obs} = \cos\alpha \cos\zeta + \sin\alpha \sin\zeta \cos\phi.
\end{equation}
Using the law of sines, the position angle $\angle MPR=\psi$ is:
\begin{equation}
  \frac{\sin\alpha}{\sin\psi} = \frac{\sin\theta_{\rm obs}}{\sin\phi}.
\end{equation}
Therefore, the sine of the position angle is
\begin{equation}
  \sin\psi = \sin\alpha \frac{\sin\phi}{\sin\theta_{\rm obs}}.
\end{equation}
Here $\sin\theta_{\rm obs}$ is required. However, only $\cos\theta_{\rm obs}$ is obtained above. Using the law of cosines for sides again for $\alpha$
\begin{equation}
  \cos\alpha = \cos\zeta \cos\theta_{\rm obs} + \sin\zeta \sin\theta_{\rm obs} \cos\psi.
\end{equation}
Then the cosine of the position angle is
\begin{equation}
  \cos\psi = \frac{\cos\alpha -\cos\zeta \cos\theta_{\rm obs}}{\sin\zeta \sin\theta_{\rm obs}}.
\end{equation}
Then it is straight forward to show that
\begin{equation}
  \tan\psi \equiv \frac{\sin\psi}{\cos\psi} = \frac{\sin\alpha \sin\phi}{\cos\alpha \sin\zeta -\sin\alpha \cos\zeta \cos\phi}.
\end{equation}
This is the commonly cited formula for the position angle. In practice, $\psi$ should be replaced by $\psi-\psi_0$, $\phi$ should be replaced by $\phi-\phi_{0}$, where $\psi_0$ and $\phi_{0}$ are constants to be determined by fitting the observations (Everett \& Weisberg 2001; Johnston \& Kramer 2019).

The third angle $\angle RMP=\phi_{\rm obs}$ is the azimuthal angle of the line of sight in the magnetic frame. Similar calculations showed that
\begin{eqnarray}
  &&\sin\phi_{\rm obs} = \sin\zeta \frac{\sin\phi}{\sin\theta_{\rm obs}} \\
  &&\cos\phi_{\rm obs} = \frac{\cos\zeta -\cos\alpha \cos\theta_{\rm obs}}{\sin\alpha \sin\theta_{\rm obs}}
\end{eqnarray}
These two equations will be used during the following calculations.

\section{Detailed calculations for a dipole field}
\label{appendixB}

\begin{figure}
\centering
\includegraphics[width=0.35\textwidth]{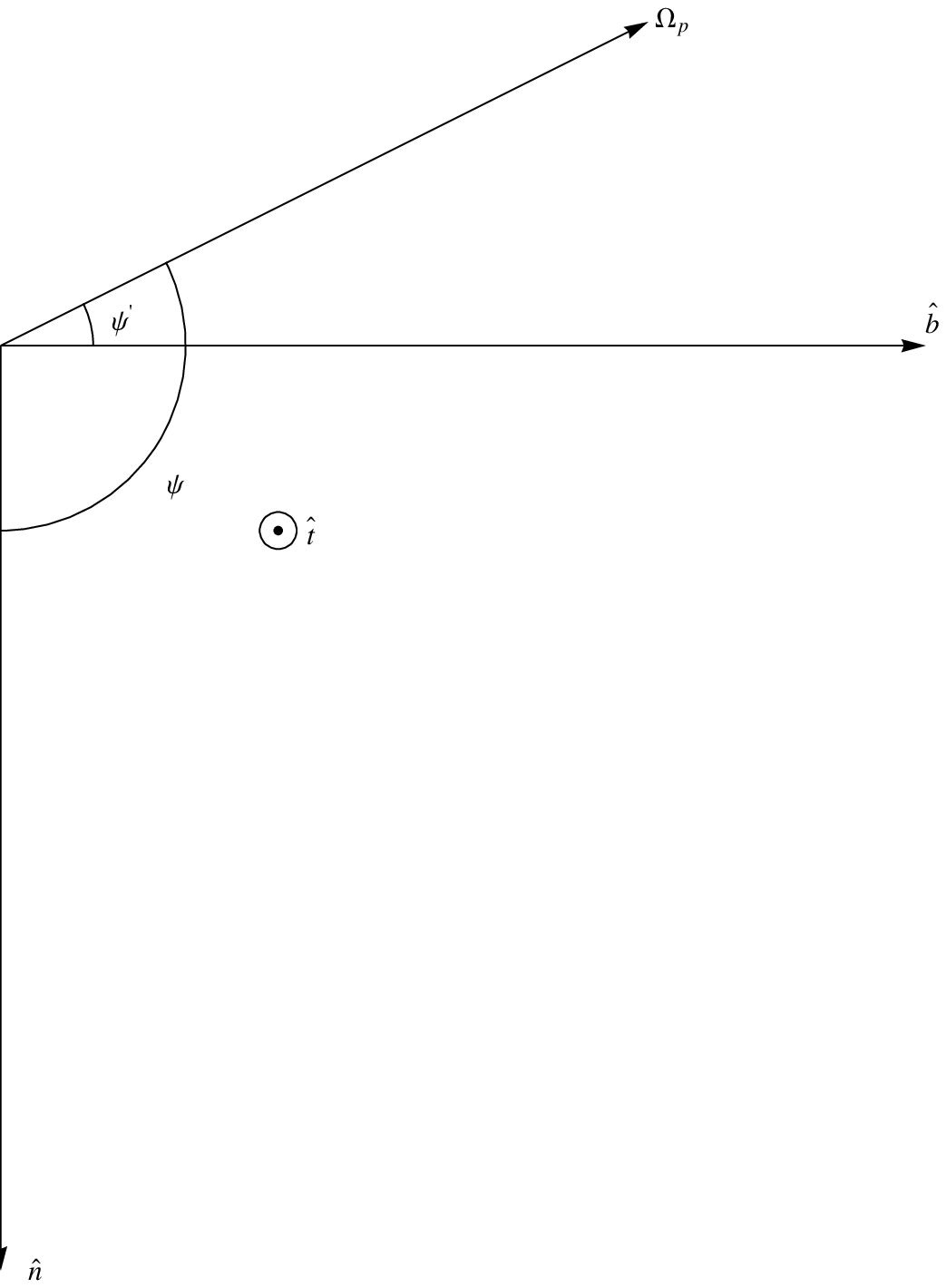}
\caption{Position angle between the normal vector and the projection of rotational axis.}
\label{fig_gbinormal}
\end{figure}

According to Hibschman \& Arons (2001), given the magnetic field, the tangent vector along the magnetic field can be found by normalization $\bf \hat{t} = B/|B|$.
The curvature vector at that point is ${\bf \kappa = (\hat{t}\cdot \nabla) \hat{t}} = 1/2 \bf \nabla \hat{t}^2 - \hat{t} \times (\nabla \times \hat{t})$. The normal vector is the normalized curveture vector $\bf \hat{n} =\kappa/|\kappa|$. The binormal vector is perpendicular to both the tangent vector and normal vector $\bf \hat{b} \equiv \hat{t} \times \hat{n}$. The projection of the rotational axis onto the plane of sky is: $\bf \Omega_{\rm p} \equiv \hat{\Omega} - (\hat{\Omega}\cdot \hat{t}) \hat{t}$, where $\bf \hat{\Omega}$ is the unit vector along the roational axis. The angle between the rotational axis and the normal vector is the position angle, shown in figure \ref{fig_gbinormal}. The angle between the rotation axis and the binormal vector is: $\cos\psi^{\prime} = \bf \hat{b} \cdot \hat{\Omega}_{\rm p}$, $\sin \psi^{\prime} = \bf \hat{t} \cdot (\hat{b} \times \hat{\Omega}_{\rm p})$. $\bf \hat{\Omega}_{\rm p}$ is the unit vector along $\bf \Omega_{\rm p}$. Therefore, the position angle is
\begin{eqnarray}
  \tan \psi &=& -\cot\psi^{\prime} = \frac{\bf \hat{b} \cdot \Omega_{\rm p}}{\bf \hat{b} \cdot (\hat{t} \times \Omega_{\rm p})}\\
  &=& \frac{\bf \hat{b} \cdot \hat{\Omega}}{\bf \hat{b} \cdot (\hat{t} \times \hat{\Omega})},
\end{eqnarray}
where during the last step further simplification are used.

For a dipole magnetic field ${\bf B_0} = (\frac{2\mu \cos\theta}{r^3}, \frac{\mu \sin\theta}{r^3}, 0)$, it can be shown that the final position angle is the same as in equation (\ref{eqn_psidipole}). The power of this detailed calculation process will be manifested when there are perturbations to the magnetic field (Hibschman \& Arons 2001).  The magnetic field may be rewritten as $\bf B= B_0 + B_1$, where $\bf B_0$ is the dipole magnetic field, $\bf B_1$ is the perturbed magnetic field. Then the tangent vector $\bf \hat{t}= \hat{t}_0 + t_1$, curvature $\bf \kappa = \kappa_0 + \kappa_1$, normal vector $\bf \hat{n} = \hat{n}_0 + n_1$, and binormal vector $\bf \hat{b} = \hat{b}_0 + b_1$ will all be perturbed. Keeping only the linear term, the perturbed quantities are
\begin{eqnarray}
  \bf t_1 &=& \bf B_1/|B_0|\\
  \bf \kappa_1 &=& \bf (\hat{t}_0 \cdot \nabla) t_1 + (t_1 \cdot \nabla) \hat{t}_0\\
  &=& \bf \nabla (\hat{t}_0 \cdot t_1) - \hat{t}_0 \times (\nabla \times t_1) - t_1 \times (\nabla \times \hat{t}_0) \\
  \bf n_1 &=& \bf \frac{1}{|\kappa_0|} (\kappa_1 - (\hat{n}_0 \cdot \kappa_1)\hat{n}_0)\\
  \bf b_1 &=& \bf \hat{t}_0 \times n_1 + t_1 \times \hat{n}_0.
\end{eqnarray}
The modification of position angle is
\begin{equation}
  \Delta \psi = \frac{\bf \hat{t}_0 \cdot (b_1 \times \Omega_{\rm p})}{\bf \hat{b}_0 \cdot \Omega_{\rm p}}.
\end{equation}

In Hibschman \& Arons (2001), they also considered the change of binormal vector due to changes in the emission point.
During this process, more assumptions are adopted (e.g., the radio emission arises at a given radius, only lateral motion is allowed).
Furthermore, different perturbations can be simply be added since only linear terms are kept. This effect (changes in emission point) and others will not affect the changes due to a perturbed magnetic field. The final result will be simple addition of different effects.
Therefore, in the main text we do not consider changes in the emission point and only consider the change of position angle due to a twisted magnetic field.

\section{The relation between emission point and  line of sight for a dipole field}
\label{appendixC}

The line of sight lies in the direction ($\theta_{\rm obs}$, $\phi_{\rm obs}$), in spherical coordinate, in the magnetic frame. In Cartesian coordinate
\begin{equation}
  \bf \hat{l} = \sin\theta_{\rm obs}\cos\phi_{\rm obs} \hat{x} + \sin\theta_{\rm obs}\sin\phi_{\rm obs} \hat{y} +
  \cos\theta_{\rm obs} \hat{z}.
\end{equation}
For a dipole field, the tangent vector at ($\theta$, $\phi$) is
\begin{eqnarray}
  {\bf \hat{t}_0} &=&\frac{1}{N_1} (\cos\theta,\frac12 \sin\theta,0)\\
  &=&  \frac{1}{N_1} (\cos\theta \hat{r} + \frac12 \sin\theta \hat{\theta}),
\end{eqnarray}
where $N_1=\sqrt{1-(3/4)\sin^2\theta}$ is the normalization factor. The relation between the base vector in spherical coordinate and
Cartesian coordinate is (Griffiths 2017)
\begin{eqnarray}
  \hat{r} &=& \sin\theta \cos\phi \hat{x} + \sin\theta \sin\phi \hat{y} + \cos\theta \hat{z}\\
  \hat{\theta} &=& \cos\theta \cos\phi \hat{x} + \cos\theta \sin\phi \hat{y} - \sin\theta \hat{z} \\
  \hat{\phi} &=& -\sin\phi \hat{x} + \cos\phi \hat{y}.
\end{eqnarray}
Then, the tangent vector rewritten Cartesian coordinate is
\begin{equation}
  {\bf \hat{t}_0} = \frac{1}{N_1} [ \frac{3}{2} \sin\theta \cos\theta \cos\phi \hat{x} + \frac{3}{2} \sin\theta \cos\theta \sin\phi \hat{y}
  + (\cos^2\theta - \frac12 \sin^2\theta) \hat{z} ]
\end{equation}
It is a general assumption that the tangent vector should be in the direction of the line of sight (Hibschman \& Arons 2001).
By equaling the tangent vector and the line of sight unit vector, it is found that
\begin{eqnarray}
  \frac{1}{N_1} \frac32 \sin\theta \cos\theta \cos\phi &=& \sin\theta_{\rm obs} \cos\phi_{\rm obs} \\
  \frac{1}{N_1} \frac32 \sin\theta \cos\theta \sin\phi &=& \sin\theta_{\rm obs} \sin\phi_{\rm obs}\\
  \frac{1}{N_1} (\cos^2\theta -\frac12 \sin^2\theta) &=& \cos\theta_{\rm obs}.
\end{eqnarray}
From the first two equations, it can be inferred that $\phi=\phi_{\rm obs}$. Then both the sine and cosine of $\theta_{\rm obs}$ is known. It can combined to obtain
\begin{equation}
  \tan\theta_{\rm obs} = \frac32 \frac{\tan\theta}{1-\frac12 \tan^2\theta}.
\end{equation}
This a quadratic equation of $\tan\theta$ and can be solved analytically. For the case of small $\theta$, it can be found directly that (Hibschman \& Arons 2001)
\begin{equation}
  \theta_{\rm obs} = \frac32 \theta.
\end{equation}

\bsp	
\label{lastpage}
\end{document}